\documentclass[12pt]{article}

\begin{document}

\noindent {\footnotesize{\it Annales de la Fondation Louis de
Broglie} {\textbf{27}}, no. 1 or 2, ... (2002), in press \\ also
arXiv.org e-print archive \ http://arXiv.org/abs/quant-ph/0202170
}

\bigskip\medskip
\begin{center}
{\LARGE \bf On the notion of the photon }
\end{center}

\vspace{2mm}

\begin{center}
{\bf Volodymyr Krasnoholovets}
\end{center}

\begin{center}
{Institute of Physics, National Academy of Sciences, \\ Prospect
Nauky 46,   UA-03028 Ky\"{\i}v, Ukraine \\ (web page \
http://inerton.cjb.net)}
\end{center}

\date{12 November 2001}

\begin{abstract}
\hspace*{\parindent} It is shown that the photon, the quantum of
electromagnetic field, allows the consideration in the framework
of the scheme which in some aspects is typical for the phonon, an
excitation of the crystal lattice of a solid. The conclusion is
drawn that the photon may be interpreted as an elementary
excitation in a fine-grained space. The corollary is in excellent
agreement with the space structure and submicroscopic quantum
mechanics, which have recently been constructed by the author in a
series of works.

\vspace{2mm}


\vspace{2mm}

general theory of fields and particles
\end{abstract}

\vspace{6mm}

Louis de Broglie was the first who proposed the detailed wave
theory of the photon [1]. In the theory, quanta of light were
composite formations: the photon regarded as a couple of Dirac
particles with very small masses. De Broglie equations decomposed
to equations for a spinless particle and the Maxwell equations
were complemented by correction terms with the electromagnetic
potentials. Thus, de Broglie's theory representing a grand
synthesis of matter and light was excellently constructed in an
ordinary space. The notions employed were very close to classical
images and yet the quantization of fields was an integral part of
the photon theory.

Today free photons and their interaction with particles are widely
described in the framework of the second quantization methodology.
However, the description of the electromagnetic field in terms of
creation and annihilation operators is justified in the wave
vector presentation. Quantum electrodynamics does not study the
problem of spatial pattern of the photon. One can only assume that
the problem is reduced to the imagination of an energetic object
that moves as a "particle-wave" (i.e. something of an
indeterminate nature) with the velocity of light $c$ and occupies
a volume $\sim \lambda^3$, where $\lambda$ is the photon
wavelength. However, on the other hand, photons obey the Bose
statistics and therefore a number of photons may collect in a
volume $v<< \lambda^3$. This is why in the case of the high
density of photons we meet with some conceptual difficulty.

One kind of such problems has been studied by the author in paper
[2]. The work deals with the investigation of the interaction of
an intensive photon pulse radiated by laser with atoms of gas and
electrons of the metal. It has been argued that a photon flux can
be regarded as a flow of corpuscles which are only several nm
apart in spite of the fact that the wavelength of photons is of
the order of several hundreds nm. Thus the examination [2] of
numerous experiments has shown that photons, indeed, might be
considered as very small corpuscles in a 3D space. But what is
spatial pattern of the photon considered from the microscopic
standpoint? To answer the question we should revise the modern
views on notions of a particle and the quantum of a field.

Bussey [3] has recently considered the phonon as a model for
elementary particles. He has analyzed "the phonon collapse" in the
one-dimensional lattice and emphasized that this is exactly
analogous to the familiar wavefunction collapse of a normal
quantum particle, such as a photon, as he notes, observed by
Mizobuchi and Ohtake [4]. Bussey notes that the atomic
displacements $u_l$ which create the phonon state resemble the
field elements $\phi (x)$ of a particle, since quantum field
theory describes elementary particles as excitations of fields
whose ground state is the vacuum. He says [3] that the existence
of such parallels between excitations of fields and phonons has
long been recognized (see, e.g. Ref. [5]).

However, one may treat quantum field theory only as an
approximation to the description of nature, just as the Fourier
approximation is applied to a concrete continuous function in
mathematics.

On the other hand, the photon might be considered as an aether
wave, as has been pointed out by Meno [6]. Nonetheless, his model
is restricted by a pure phenomenological description of a wave in
an aether treated as a gas of incompressible "gyrons".

In the recent concept by the author [7-13] particles appear as
stable local deformations of a degenerate space that is regarded
as a 3D elastic cellular net with the size of a cell $\sim
10^{-28}$ cm (recall that at this scale all types of physical
interactions come together) [12,13]. In this case we do not need
abstract field elements $\phi (x)$. Any motion of the deformation
is accompanied by excitations of the space net that were called
inertons [7] and the motion of such a complex formation, i.e. a
particle surrounded by its inerton cloud, falls within  the
formalism of quantum mechanics both Schr\"odinger's [7,8] and
Dirac's [9]. In paper [10] we have theoretically studied the
collective motion of atoms, i.e. phonons, in the crystal lattice
assuming the existence of inerton clouds in their surrounding and
then proved it experimentally (other manifestations of clouds of
inertons have been demonstrated in Ref. [2]). Thus owing to a
great success of the new concept we may assume that photons
migrated in the space net might be considered as something similar
to phonons which are excited in the crystal lattice.

Phonons appear due to spontaneous vibrations of atoms in the
crystal lattice. Actually owing to the atom-atom interaction, one
can write the Lagrangian
\begin{equation}\label{1}
L=\frac 12 \sum_{\vec n} m {\kern 1pt}{\dot{\vec r}}_{\vec
n}^{\kern 3pt {2}} - \frac 12 {\sum_{\vec n,{\kern 1pt} \vec
{n^\prime}}}^{\prime} \gamma_{\vec n {\kern 1pt} \vec {n^\prime}}
{\kern 1pt}{\vec r}_ {\vec n}{\kern 1pt}{\vec r}_{\vec {n^\prime}}
\end{equation}
where $m$ is the mass of the $\vec n$th atom, ${\vec r}_{\vec n}$
and  ${\dot {\vec r}}_{\vec n}$ are the displacement of the atom
from the equilibrium position and the velocity of the atom
respectively, \ $\gamma_{\vec n {\kern 1pt} \vec {n^\prime}}$ is
the tensor of elasticity interaction of atoms, the dot over the
vector ${\dot{ \vec r}}_{\vec n}$ means the differentiation with
respect to the proper time of the crystal (compare with Ref. [8]).
As is well known, expression (1) can be rewritten via the
generalized, or canonical, coordinates $A_{\vec k s}$ and $ \dot
A_{\vec k s} \equiv P_{\vec k  s}$ (see, e.g. Ref. [14])
\begin{equation}\label{2}
L=\frac 12 \sum_{\vec k, {\kern 2pt} s}\Big( \dot A_{\vec k s}
\dot A_{-\vec k s}
  - \Omega^2_s(\vec k) A_{\vec k s} A_{-\vec k s} \Big)
\end{equation}
where $\Omega^2_s(\vec k)$ is the frequency of the $s$th branch of
acoustic vibrations of atoms. $\dot A_{-\vec k s}$ denotes the
generalized momentum $P_{\vec k s}$. Coordinates $A_{\vec k s}$
and $P_{\vec k s}$ are substituted for the corresponding operator
\begin{eqnarray}
&& A_{\vec k s} \rightarrow \hat A_{\vec k s} = \sqrt
{\hbar/2\Omega_s(\vec k)} \  (\hat b_{\vec k s} + \hat
b^\dag_{-\vec k s});   \\     \nonumber &&P_{\vec k s} \rightarrow
\hat P_{\vec k s} = i \sqrt{\hbar\Omega_s(\vec k)/2} \ (\hat
b^\dag_{\vec k s} -\hat b_{-\vec k s}). \label{3}
\end{eqnarray}
Here, $\hat b^\dag_{\vec k s}(\hat b_{\vec k s})$ is the Bose
operator of creation (annihilation) of a phonon. In terms of the
second quantization operators the energy operator of the lattice
vibrations takes the form
\begin{equation}\label{4}
\hat H = \sum_{\vec k, {\kern 2pt} s} \hbar \Omega_s(\vec k) \Big(
\hat b^\dag_{\vec k s}\hat b_{\vec k s} + \frac 12 \Big).
\end{equation}

   Now let us proceed to the inspection of the energy operator of
a free electromagnetic field, which has the same form
\begin{equation}\label{5}
\hat {\cal H} = \sum_{\vec k, {\kern 2pt} s} \hbar \omega_s(\vec
k) \Big( \hat a^\dag_{\vec k s}\hat a_{\vec k s} + \frac 12 \Big)
\end{equation}
where $\hat a^\dag_{\vec k s}(\hat a_{\vec k s})$ is the Bose
operator of creation (annihilation) of a photon, an elementary
excitation of the electromagnetic field; $\omega_s(\vec k)$ is the
cyclic frequency of the photon with the wave vector $\vec k$ and
the $s$ polarization. In spite of the similarity of expressions
(4) and (5), their original classical Lagrangians are absolutely
different. In the case of phonons we start from the discrete
function (1), but in the case of photons we emanate from the
Lagrangian density  of the continual electromagnetic field
\begin{equation}\label{6}
 {\cal L} = \frac {1}{8 \pi} \Big\{  \frac 1{c^{\kern 1pt 2}}
 \Big( \frac{\partial \vec A}{\partial {\kern 1pt} t} \Big)^2
 - (\nabla \times \vec A)^2  \Big\}
\end{equation}
where $\vec A$ is the vector potential of the field. However if we
formally do an analysis trying to advance from the energy operator
(5) to an initial classical Lagrangian keeping the phonon scheme
above, we will come to a very interesting finding.

First, the operators of canonical variables expressed in terms of
${\hat a}_{\vec k s}^\dag$ and ${\hat a}_{\vec k s}$ are
\begin{eqnarray}
&&{\hat {\cal A}}_{\vec k s}(t) = \sqrt {2\pi \hbar c^{{\kern 1pt}
2} /\omega_{\vec k}} \ \big( \hat a_{\vec k s}(t) + \hat
a^\dag_{-\vec k s}(t) \big);
\\ \nonumber &&{\hat{\cal P}}_{\vec k s}(t) =
i \sqrt{\hbar \omega_{\vec k}/8\pi c^{{\kern 1pt} 2}}\ \big( \hat
a^\dag_{\vec k s}(t) -\hat a_{-\vec k s}(t) \big). \label{7}
\end{eqnarray}
Then the corresponding canonical variables are
\begin{eqnarray}
&&{\vec {\cal A}}(\vec r, t) = \frac 1{\sqrt V} \sum_{\vec k,
{\kern 2pt} s} \vec e_{\vec k s} {\cal A}_{\vec k s}(t)e^{{\kern
1pt}i\vec k \vec r};
\\          \nonumber &&{\vec {\cal P}}(\vec r, t)
= \frac 1{\sqrt V} \sum_{\vec k, {\kern 2pt} s} \vec e_{\vec k
s}{\cal P}_{\vec k s}(t) e^{- i \vec k \vec  r}. \label{8}
\end{eqnarray}
Formulas (8) present the Fourier-series expansion of the classical
variables $\vec {\cal A}(\vec r, t)$  and $\vec {\cal P}(\vec r,
t)\equiv (1/4\pi c^{{\kern 1pt} 2}){\kern 0.7 pt}\partial \vec
{\cal A}/\partial {\kern 1pt}t$ \ in the volume $V$. And the
expansion, indeed, adequately depicts the actual discrete
structure of electromagnetic field, i.e., photonic nature of the
field. Thus one can rewrite the Lagrangian density (6) as
\begin{equation}\label{9}
 {\cal L} = \frac {1}{8 \pi V} \sum_{\vec k, {\kern 2pt} s}  \Big[ \frac 1
 {c^{{\kern 1pt}2}}
  {\kern 1pt}\frac{\partial {\cal A}_{\vec k s}}{\partial{\kern 1pt}t}
 {\kern 1pt}\frac{\partial {\cal A}_{-\vec k s}}{\partial{\kern 1pt} t}
 - \big( \nabla \times \vec e_{\vec k s}{\cal A}_{\vec k s} \big)
 \big( \nabla \times \vec e_{\vec k s}{\cal A}_{-\vec k s} \big)
 \Big].
 \end{equation}
Let us invert expression (9) written via the wave vector
presentation to that written in terms of discrete spatial
variables
\begin{equation}\label{10}
{\cal L} = \frac {1}{8 \pi V} \sum_{\vec n}  \Big[ \frac
1{c^{{\kern 1pt}2}} {\kern 1pt}\Big( \frac{\partial \vec {\cal
A}_{\vec n}}{\partial {\kern 1pt}t} \Big)^2  - \big( \nabla_{\vec
n} \times {\vec {\cal A}}_{\vec n} \big)^2  \Big].
 \end{equation}
Under this new description, the vector $\vec n$ plays the role of
the radius vector that defines knots of a lattice of space or
cells of a space net. The most credible speculation is that the
space consists of cells (or balls, or superparticles) which are
closely packed [7-9,11-13]. Then the equation of motion
\begin{equation}\label{11}
\frac{\partial^{{\kern 1pt}2} \vec {\cal A}_{\vec
n}}{\partial{\kern 1pt}t^{{\kern 1pt}2}} - \frac 1{c^{{\kern
1pt}2}} \nabla_{\vec n}^{{\kern 1pt}2} {\kern 1pt}{\vec {\cal
A}}_{\vec n}=0
\end{equation}
followed by expression (10) specifies the behavior of some kind of
a polarization $\vec {\cal A}_{\vec n}$ localized in the cell
which position in the space net is defined by the $\vec n$th
radius vector in the moment $t$.

From the results obtained it may be deduced that the photon should
be regarded as an elementary excitation, or quasi-particle of some
sort that migrates hopping from cell to cell in the space net
rather than a canonical particle-wave of an undetermined nature
that moves in an empty space, or a dim vacuum. In this case the
vector potential $\vec {\cal A}_{{\kern 0.5pt}\vec n}$ should be
interpreted as some kind of polarization/deformation that is
induced in an incoming cell along a path of the corresponding
photonic excitation. The vector of local peculiar deformation
$\vec {\cal A}_{\vec n}$ changes in any point (i.e. cell) that is
characterized by the radius vector $\vec n$ according to Eq. (11).
But the migration of the photon "core" features the equation
\begin{equation}\label{12}
\frac{d}{d {\kern 1pt} t}{\kern 1pt}{\vec n}= \frac {\vec n}{n}
{\kern 1pt} c
\end{equation}
where the proper time $t >\tau$ and $\tau$ is the "inoculating"
time (i.e. photon lifetime in a cell, see below). Thus, a couple
of equations (11) and (12) completely describe the behavior of the
photon.

Consequently, we may conclude that unlike the phonon that
envelopes a great number of the lattice sites (the phonon is
enclosed in a volume $\sim k_{{\kern 1pt}\rm ac}^{-3}$ where
$k_{{\kern 1pt}\rm ac}=2\pi/\lambda_{{\kern 1pt}\rm ac}$ is the
wave number and $\lambda_{{\kern 1pt}\rm ac}$ is the wavelength of
the appropriate acoustic excitation), the photon may be considered
as something that is located only in one cell of the space.
However, the lifetime of the photon in one cell whose size $a \sim
10^{-28}$ cm  should be extremely short. It can be estimated if we
divide the period $T$ of the photon by a number $N$ of cells that
cover the section of spatial period $\lambda =2\pi/ k$ of the
photon. Therefore $N=\lambda /a$ and hence the lifetime $\tau =T
a/\lambda$. For instance in the case of an optical photon $\lambda
\sim 10^{-8}$ cm, $N \sim 10^{36}$ and then $T \sim 1$ fs and
$\tau \sim 10^{-35}$ s.

The notion of the wavelength $\lambda$ of the photon specifies the
spatial period at which the cell's polarization restores its
initial state. In other words, the photon wavelength $\lambda$ is
a distance which the photon should run hopping from cell to cell
in order that the initial phase of the photon polarization be
restored. The period of polarization oscillations $T=1/\nu$ is
connected with $\lambda$ by relation $c=\lambda/T$. Yet the real
size of the photon, i.e. size of its "core", is limited by the
size of a cell of the space net $a \sim 10^{-28}$ cm.

Thereby, the photon that is characterized by the energy
$\varepsilon = h \nu$ makes up an excitation of the space net,
which migrates by structural cells of the net and carriers a
polarization from cell to cell. This hopping motion is similar to
the migration of Frenkel, or molecular excitons in molecular
crystals.

The interaction of the photon with a matter is realized by means
of the interaction of the photon with inerton clouds of the matter
entities such as electrons, atoms, etc. as the radius of the dense
inerton cloud that surrounds an electron or an atom much exceeds
the mean distance between the entities in condensed media [2,10].
One can raise the question, how can inertons which represent a
substructure of the matter waves of any charged particle be
distinguished from the particle's virtual photons which feature
the particle's electromagnetic field? Inertons were uniquely
determined [7,8,12,13] as elementary excitations, or
quasi-particles of the space net, which carry a local deformation
of the net. In order for agreement the particle's electrodynamics
to particle's quantum mechanics, photons must be treated as the
same quasi-particles, i.e. massive inertons, which however are
endowed with an additional property. Thereby the notion of
electromagnetic "polarization" of a cell, i.e. a special
additional deformation of a cell, should be clarified and this
indeed is quite possible in the framework of the model that is
developing.

A theory of the photon-inerton interaction being constructed will
allow concrete practical applications. Namely, it will make
possible: (i) the disclosing a microscopic mechanism of the
phenomenon of the diffraction of light, which so far is still
described in pure geometrical (i.e. phenomenological) terms; (ii)
a microscopic description of the phenomenon of the bending of a
luminous ray in the vicinity of a star, which is till now
considered solely in the framework of the general relativity
macroscopic approach; (iii) a detailed theoretical analysis of
entangled states; (iv) the construction of submicroscopic models
describing the photon-photon cross-section into hadrons (modern
models, see e.g. Refs. [15,16], operate with $\gamma$-photons as
very complicated formations. We may assume that those
$\gamma$-photons are not canonical photons described above but
rather peculiar cooperative excitations of a sort. In fact, the
concept is not contradictory to elementary particle physics: the
photon is not included in the list of {\it elementary particles}
[17]).

Thus, the results presented in this work permits the tracing a
microscopic physical pattern of the photon in a 3D space, or 4D
space-time. The new concept of the photon turns the study of the
photon properties to an inner structure of a cell, or
superparticle, the building block of the real space.

\end{document}